\documentclass{aa}
\usepackage{graphicx}
\usepackage{txfonts}

\def\simlt{\lower.5ex\hbox{$\; \buildrel < \over \sim \;$}}
\def\simgt{\lower.5ex\hbox{$\; \buildrel > \over \sim \;$}}

\begin{document}

\title{Cyclical period changes in HT Cas: a clear difference between
  systems above and below the period gap\thanks{Based on observations
    made at the Astronomical Station Kryoneri, owned by the National
    Observatory of Athens, Greece.}}

\author{B. W. Borges 
  \inst{1}
  \and R. Baptista
  \inst{1}
  \and C. Papadimitriou
  \inst{2}
  \and O. Giannakis
  \inst{2}
}

\offprints{B. W. Borges}

\institute{
  Departamento de F\'{\i}sica,
  Universidade Federal de Santa Catarina, CEP 88040-900,
  Florian\'{o}polis, Brazil \\
  \email{bernardo@astro.ufsc.br, bap@astro.ufsc.br}
  \and
  Institute of Astronomy and Astrophysics, 
  National Observatory of Athens, PO Box 20048,
  Athens 11810, Greece \\
  \email{cpap@astro.noa.gr, og@astro.noa.gr}
}

\date{Received ?? August 2007 / Accepted ?? August 2007}

\abstract
{}
{We report the identification of cyclical changes in the orbital
  period of the eclipsing cataclysmic variable HT~Cas.}
{We measured new white dwarf mid-eclipse timings and combined with
  published measurements to construct an observed-minus-calculated
  diagram covering 29 years of observations.}
{The data present cyclical variations that can be fitted by a linear
  plus sinusoidal function with period $36$~yr and semi-amplitude
  $\sim 40$~s.  The statistical significance of this period by an
  F-test is larger than 99.9 per cent.}
{We combine our results with those in the literature to revisit the
  issue of cyclical period changes in cataclysmic variables and their
  interpretation in terms of a solar-type magnetic activity cycle in
  the secondary star. A diagram of fractional period change ($\Delta
  P/P$) versus the angular velocity of the active star ($\Omega$) for
  cataclysmic variables, RS~CVn, W~UMa and Algols reveal that close
  binaries with periods above the gap (secondaries with convective
  envelopes) satisfy a relationship $\Delta P/P \propto \Omega^{-0.7
    \pm 0.1}$.  Cataclysmic variables below the period gap (with fully
  convective secondaries) deviate from this relationship by more than
  3-$\sigma$, with average fractional period changes $\simeq 5$ times
  smaller than those of the systems above the gap.}


\titlerunning{Cyclical period changes in HT Cas}

\maketitle

\section{Introduction} 
\label{sec:intr}

HT Cassiopeiae (HT~Cas) is a short-period ($P_{{\rm orb}} = 1.77$ hr)
eclipsing cataclysmic variable (CV). In these binaries, a late-type
star (the secondary) overfills its Roche lobe and transfers matter to
a companion white dwarf (Warner 1995). The evolution of CVs is mainly
driven by two ingredients: angular momentum loss, to sustain the mass
transfer process, and the response of the secondary to the mass loss.
Two mechanisms were proposed for angular momentum loss in CVs. The
first is gravitational radiation, which is effective only for short
orbital periods (Patterson 1984). The second is angular momentum
carried away by a stellar wind magnetically coupled to the secondary
surface (magnetic braking mechanism; Rappaport et~al. 1983; King
1988). As CVs are tidally locked binaries, any momentum lost by the
secondary is also subtracted from the total orbital momentum of the
system, causing a secular decrease of the orbital period.

In the disrupted braking model of CV evolution (Rappaport et~al.
1983; Hameury et~al. 1991), the observed dearth of systems with
periods in the range 2-3~hr (known as ``period gap''; Knigge 2006) is
explained by a sudden drop in the efficiency of the magnetic braking
mechanism when the secondary, evolving from longer orbital periods,
reach $P_{{\rm orb}} \sim 3$~hr and becomes fully convective ($M_{2}
\sim 0.3\,M_{\odot}$). However, there are some key assumptions in the
standard model, most notably concerning angular momentum loss, that
are inconsistent with both the observed spin-down of young, low-mass
stars and theoretical developments in our understanding of stellar
winds (Andronov et~al.  2003).

The secular evolution of the binary can in principle be detected by
measuring the changes in the orbital period of eclipsing CVs.
Eclipses provide a fiducial mark in time and can usually be used to
determine the orbital period (and its derivative) with high precision.
However, attempts to measure the long-term orbital period decrease in
CVs have been disappointing: none of the studied stars show the
expected rate of orbital period decrease. Instead, most of the well
observed eclipsing CVs\footnote{i.e., those with well-sampled
  observed-minus-calculated (O$-$C) eclipse timings diagram covering
  more than a decade of observations.} show cyclical period changes
(Baptista et~al. 2003 and references therein). Cyclical orbital period
variations are also observed in other close binaries with late-type
components -- Algols, RS CVn and W UMa systems (Lanza \& Rodon\`{o}
1999). The most promising explanation of this effect seems to be the
existence of a solar-type (quasi- and/or multi-periodic) magnetic
activity cycle in the secondary star. A number of mechanisms have been
proposed which are capable of producing a modulation of the orbital
period on time scales of decades, induced by a variable magnetic field
in the convective zone of the late-type component (Matese \& Whitmire
1983; Applegate \& Patterson 1987; Warner 1988; Applegate 1992;
Richman et~al. 1994; Lanza et~al. 1998; Lanza 2006a). The relatively
large amplitude of these cyclical period changes probably contributes
to mask the low amplitude, secular period decrease.

This paper reports the results of an investigation of orbital period
changes in HT~Cas. The observations and data analysis are described in
section \ref{sec:observ}. A discussion of cyclical orbital period
changes in CVs, above and below the period gap, are presented in
section \ref{sec:discuss}.

\section{Observations and data analysis}
\label{sec:observ}

Time-series of white light CCD photometry of HT~Cas were obtained
during 5 nights on 2007 January/February with the 1.2-m telescope at
the Astronomical Station Kryoneri (Greece). The data cover a total of
11 eclipses and were obtained with a SI-502 CCD array with $516 \times
516$ pixels. All observations have a time resolution of 25~s. A
summary of these observations is given in Table~\ref{tab:log}. The CCD
data reductions were done with IRAF\footnote{IRAF is distributed by
  the National Optical Astronomy Observatory, which is operated by the
  Association of Universities for Research in Astronomy Inc., under
  contract with the National Science Foundation.} routines and
included bias and flat-field corrections. Aperture photometry was
carried out with the APPHOT package. Time-series were constructed by
computing the magnitude difference between the variable and a
reference comparison star. HT~Cas was 0.5~mag fainter on Jan 18 and 20
with respect to the data of the remaining nights. This behavior is
reminiscent of that previously seen by Robertson \& Honeycutt (1996),
who found that HT~Cas switches between high and low brightness states
differing by 1.3~mag on time-scales from days to months. The spread in
out-of-eclipse flux within either the low or high states is much
smaller than the 0.5-mag transition occurred between Jan 20 and 21.
This systematic flux difference lead us to group the eclipse light
curves per brightness state (low and high) for the determination of
the mid-eclipse timings.

%
\begin{table}
  \caption{Log of the observations}
  \begin{tabular}{lcccc}
    \hline
    Date   & UT time & Cycles  &  Night & Quiesc. \\
    (2007) &         &         & quality$^{\ast}$ & state  \\
    \hline
    Jan 18 & 17:27-19:17 & 141096, 141097 & b & low   \\ 
    Jan 20 & 16:45-21:15 & 141123, 141124 & a & low   \\
    Jan 21 & 17:38-20:53 & 141137, 141138 & b & high  \\
    Jan 22 & 16:48-23:03 & 141150-141153  & c & high  \\
    Feb 15 & 17:15-19:15 & 141476         & b & high  \\
    \hline
    \multicolumn{5}{l}{\footnotesize{$^{\ast}$Quality: a - photometric,
        b - good, c - poor.}}
  \end{tabular}
  \label{tab:log}
\end{table}

Mid-eclipse times were measured from the mid-ingress and mid-egress
times of the white dwarf eclipse using the derivative technique
described by Wood et~al. (1985). For each brightness state, the light
curves were phase-folded according to a test ephemeris and sorted in
phase to produce a combined light curve with increased phase
resolution. The combined light curve is smoothed with a median filter
and its numerical derivative is calculated. A median-filtered version
of the derivative curve is then analyzed by an algorithm which
identifies the points of extrema (the mid-ingress/egress phases of the
white dwarf). The mid-eclipse phase, $\phi_0$, is the mean of the two
measured phases. For both data sets the difference between the
measured mid-egress and mid-ingress phases is consistent with the
expected width of the white dwarf eclipse, $\Delta\phi= 0.0493 \pm
0.0007$~cycles (Horne et~al. 1991). Finally, we adopt a cycle number
representative of the ensemble of light curves and compute the
corresponding observed mid-eclipse time (HJD) for this cycle including
the measured value of $\phi_0$. This yields a single, robust
mid-eclipse timing estimate from a sample of eclipse light curves.
These measurements have a typical accuracy of $\simeq$~5~s. The
inferred HJD timings for the representative cycles of the low- ($E =
141110$) and high-state ($E = 141194$) sets are $2,454,120.29368(5)$
and $2,454,126.48001(5)$, respectively (the uncertainties are given in
parenthesis).

Feline et~al. (2005) added new optical timings from high-speed
photometry to those listed by Patterson (1981), Zhang et~al. (1986)
and Horne et~al. (1991) to derive a revised linear ephemeris for
HT~Cas. These authors do not report evidence of period decrease or
modulation, perhaps because their analysis does not include the
optical timings of Wood et~al. (1995) and Ioannou et~al. (1999). In
the present work, the set of timings used includes mid-eclipse timings
measured from our light curves and all mid-eclipse timings from the
literature\footnote{Only timings while HT~Cas was in quiescence were
  considered, including an X-ray timing obtained by Mukai et~al.
  (1997).}. It covers a time interval of 29 yr, from 1978 to 2007.
For HT~Cas the difference between universal time (UT) and terrestrial
dynamical time (TDT) scales amounts to 26~s over the data set. The
amplitude of the difference between the barycentric and the
heliocentric correction is about 4~s. All mid-eclipse timings have
been corrected to the solar system barycentre dynamical time (BJDD),
according to the code by Stumpff (1980). The terrestrial dynamical
(TDT) and ephemeris (ET) time scales were assumed to form a contiguous
scale for our purposes.

\begin{table}
  \caption{Average mid-eclipse timings}
  \begin{tabular}{crccl}
    \hline
    Year & Cycle  &    BJDD      & (O$-$C)$^{\dagger}$  & Ref.\\
         &        & (+2400000 d) & ($\times 10^{-5}$d)  &     \\ 
    \hline
     1978 &   1076 & 43807.18220(3) & $-23$ & 1  \\
     1982 &  20824 & 45261.56732(6) & $-7$  & 2,3\\
     1983 &  25976 & 45640.99769(6) & $-9$  & 2,3\\
     1984 &  30766 & 45993.76780(5) & $-8$  & 2  \\
     1991 &  63228 & 48384.50361(5) & $+23$ & 4  \\
     1994 &  79770$^{\ddag}$ & 49602.77584(5) & $+43$ & 5 \\
     1995 &  85616 & 50033.31752(4) & $+56$ & 6  \\
     1997 &  92628 & 50549.73159(2) & $+44$ & 6  \\
     2002 & 119542 & 52531.87205(3) & $+8$  & 7  \\
     2003 & 125130 & 52943.41261(4) & $+8$  & 7  \\
     2007 & 141152 & 54123.38773(3) & $-29$ & This work.\\
    \hline
    \multicolumn{5}{l}{$\ ^{\dagger}$O$-$C times with respect to the
      linear ephemeris of Table~\ref{tab:ephem}.} \\
    \multicolumn{5}{l}{$\ ^{\ddag}$X-ray mid-eclipse timing.} \\
    \multicolumn{5}{l}{\emph{References}: (1) Patterson 1981; (2)
      Zhang et~al. 1986;} \\
    \multicolumn{5}{l}{(3) Horne et~al. 1991; (4) Wood et~al. 1995;
      (5) Mukai} \\
    \multicolumn{5}{l}{ et~al. 1997; (6) Ioannou et~al. 1999; (7)
      Feline et~al. 2005.}
  \end{tabular}
  \label{tab:timing}
\end{table}

Observed-minus-calculated times with respect to a test ephemeris were
evaluated for each timing in our data set. For a given year, annual
average values of (O$-$C) were computed for a representative cycle
number.  Finally, the average mid-eclipse timing (in BJDD)
corresponding to the representative cycle are obtained by adding the
average (O$-$C) value to the mid-eclipse time predicted by the test
ephemeris. The uncertainties were assumed to be the standard deviation
of each annual timing set. The average mid-eclipse timings are listed
in Table~\ref{tab:timing}. The corresponding uncertainties in the last
digit are indicated in parenthesis. The data points were weighted by
the inverse of the squares of the uncertainties in the mid-eclipse
times. Table~\ref{tab:ephem} presents the parameters of the best-fit
linear, quadratic and linear plus sinusoidal ephemerides with their
1-$\sigma$ formal errors quoted. We also list the root-mean-square
values of the residuals, $\sigma$, and the $\chi^{2}_{\nu}$ value for
each case, where $\nu$ is the number of degrees of freedom. In order
to check the sensitivity of the results to the uncertainty of the
timings, we repeated each fit assuming equal errors of $5 \times
10^{-5}$~d to the data points. The parameters obtained this way are
equivalent to those given in Table~\ref{tab:timing} within the
uncertainties.

\begin{table}
  \caption{Ephemerides of HT Cas}
  \begin{tabular}{ll}
    \hline
    \multicolumn{2}{l}{\bf Linear ephemeris:} \\
    \multicolumn{2}{l}{BJDD = T$_{0}$ + P$_{0}\cdot E$} \\ [1ex]
    T$_{0} = 2443727.93804\,(\pm 3)$ d & P$_{0} = 0.0736472029\,(\pm 3)$ d \\
    $\chi^{2}_{\nu_{1}}= 56.9,
    \;\;\; \nu_{1}$ = 9 & $\sigma_{1}= 29.0 \times 10^{-5}$ d \\ [2ex]
    
    \multicolumn{2}{l}{\bf Quadratic ephemeris:} \\
    \multicolumn{2}{l}{BJDD = T$_{0}$ + P$_{0}\cdot E$ + $c\cdot E^{2}$} \\ [1ex]
    T$_{0} = 2443727.93768\,(\pm 3)$ d & P$_{0} = 0.0736472230\,(\pm 9)$ d \\
    c $ = (-137 \pm 7) \times 10^{-14}$ d & $ \sigma_{2} = 11.8 \times 10^{-5}$ d \\
    $\chi^{2}_{\nu_{2}}= 9.47,
    \;\;\; \nu_{2}$ = 8 & \\ [2ex]
    
    \multicolumn{2}{l}{\bf Sinusoidal ephemeris:} \\
    \multicolumn{2}{l}{BJDD = T$_{0}$ +
      P$_{0}\cdot E$ + A$\cdot \cos\,[2\pi (E-{\rm B})/{\rm C}]$} \\ [1ex]
    T$_{0}= 2443727.93828\,(\pm 6)$ d & B $= (89 \pm 4) \times 10^{3}$ cycles \\
    P$_{0}= 0.073647200\,(\pm 2)$ d   & C $= (180 \pm 20) \times 10^{3}$ cycles \\
    A $= (46 \pm 6) \times 10^{-5}$ d & $\sigma_{\rm S}= 6.79 \times 10^{-5}$ d \\
    $\chi^{2}_{\nu_{\rm S}}= 3.11,
    \;\;\; \nu_{\rm S}= 6$ & \\
    \hline
  \end{tabular}
  \label{tab:ephem}
\end{table}

\begin{figure}
  \includegraphics[scale=0.47]{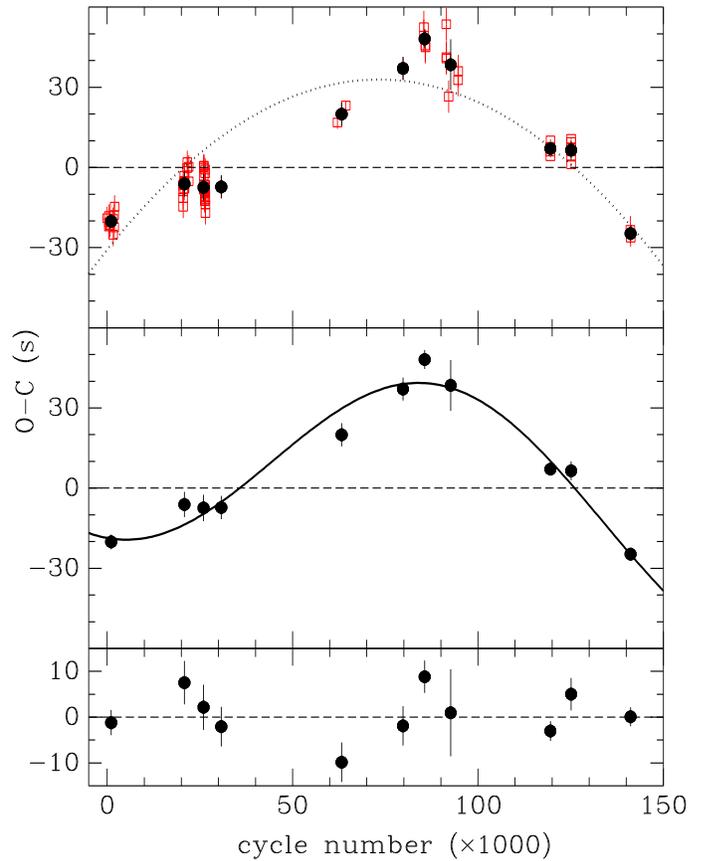}
  \caption{The O$-$C diagram of HT~Cas with respect to the linear
    ephemeris of Table~\ref{tab:ephem}. The individual timings from
    the literature and eclipse timings measured from our light curves
    are shown as open squares, while the average timings of
    Table~\ref{tab:timing} are denoted by solid circles. The dotted
    line in the upper panel depicts the best-fit quadratic ephemeris
    while the solid line in the middle panel shows the best-fit linear
    plus sinusoidal ephemeris of Table~\ref{tab:ephem}. The lower
    panel displays the residuals with respect to the linear plus
    sinusoidal ephemeris of the middle panel.}
  \label{fig:diagoc}
\end{figure}

Fig.~\ref{fig:diagoc} presents the (O$-$C) diagram with respect to the
linear ephemeris of Table~\ref{tab:ephem}. The annual average timings
of Table~\ref{tab:timing} are indicated by solid circles and show a
clear modulation. Open squares show the individual mid-eclipse timings
taken from the literature (see references in Table~\ref{tab:timing})
and individual eclipse timings measured from our light curves. The
significance of adding additional terms to the linear ephemeris was
estimated with the F-test, following the prescription of Pringle
(1975). The quadratic ephemeris has a statistical significance of 96.7
per cent with $F(1,9) = 15.1$. On the other hand, the statistical
significance of the linear plus sinusoidal ephemeris with respect to
the linear fit is larger than 99.95 per cent, with $F(3,9) = 86.2$.
The best-fit period of the modulation in HT~Cas is $36 \pm 4$~yr. The
best-fit linear plus sinusoidal ephemeris is shown as a solid line in
the middle panel of Fig.~\ref{fig:diagoc}, while the residuals with
respect to this ephemeris are shown in the lower panel.

A search for variable cycle period or for harmonics of the main cycle
period (by performing separated fits to different parts of the data
set) is not conclusive in this case because of the relatively short
time span of the data in comparison to the cycle period. However, the
eclipse timings show systematic and significant deviations from the
best-fit linear plus sinusoidal ephemeris. The fact that
$\chi^{2}_{\nu} > 1$ emphasizes that the linear plus sinusoidal
ephemeris is not a complete description of the data, perhaps signaling
that the period variation is not sinusoidal or not strictly periodic.

\section{Discussion}
\label{sec:discuss}

Our results reveals that the orbital period of HT~Cas shows
conspicuous period changes of semi-amplitude $\sim$~40~s which seems
to repeat on a time-scale of about 36~yr. The present work increases
the sample of eclipsing CVs in which orbital period modulations were
observed and motivated us to update the comparison of cyclical period
changes of CVs above and below the period gap performed by Baptista
et~al. (2003).

\subsection{Orbital period modulations in CVs}
\label{sub:mod}

This section reviews the current observational picture on the
detection of cyclical period changes in eclipsing CVs. We first
address the observational requirements needed to allow detection of
cyclical period modulations and then discuss the observational
scenario which emerges when a complete sample is constructed based on
these requirements. Cyclical orbital period changes are seen in many
eclipsing CVs (see Baptista et~al. 2003). The cycle periods range from
5~yr in IP~Peg (Wolf et~al 1993) to about 36~yr in HT~Cas, whereas the
amplitudes are in the range $10-10^{2}$~s. 

\begin{figure}
  \includegraphics[scale=0.47]{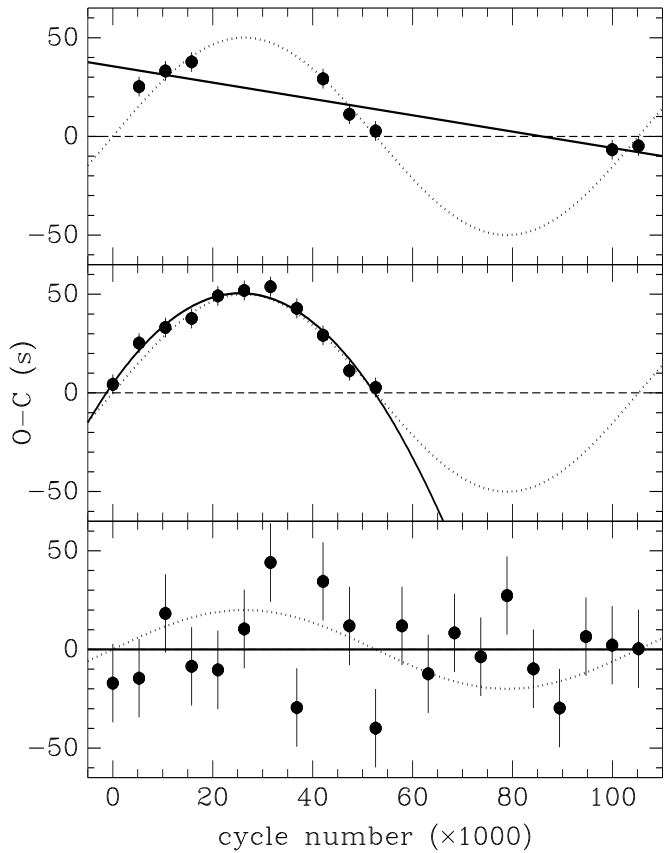}
  \caption{Influence of data sampling (upper panel), time coverage 
    (middle panel) and accuracy of the eclipse timings (lower panel)
    in the detection of orbital period modulations. The dotted line
    represents (O$-$C) values constructed from a period modulation of
    20~yr and amplitude 50~s (20~s in the lower panel). Gaussian noise
    of amplitudes 5~s (upper and middle panels) and 20~s (lower panel)
    were added to the (annually sampled) timings to simulate the
    uncertainties of a real data set. The solid line depicts the
    best-fit ephemeris obtained for the synthetic data used (solid
    circles) to illustrate each case.}
  \label{fig:simul}
\end{figure}

A successful detection of these cyclical period changes demands an
(O$-$C) diagram covering at least one cycle of the modulation (i.e.,
at least about a decade of observations) and that the uncertainty in
the (annually averaged) eclipse timings is smaller than the amplitude
of the period modulation to allow a clean detection of the latter.
Therefore, decade-long time coverage and high precision eclipse
timings (better than 10~s and 20~s, respectively for systems below and
above the period gap) are basic requirements. A third key aspect
concerns the time sampling of the observations. An (O$-$C) diagram
constructed from sparse and infrequent eclipse timing measurements may
easily fail to reveal a cyclical period change. Figure~\ref{fig:simul}
illustrates this argument. It shows synthetic (O$-$C) diagrams
constructed from a period modulation of 20~yr and amplitude 50~s
(dotted line). Gaussian noise of amplitude 5~s was added to the annual
timings to simulate the typical uncertainties of a real data set. The
best-fit ephemeris is indicated by a solid curve/line in each case.
The upper panel shows the case of poor data sampling. The gaps around
eclipse cycles $\sim 30000$ and $60000-100000$ mask the period
modulation and the avaliable data (solid circles) is best fit by a
linear ephemeris (solid line). HT~Cas itself is a good example of the
poor sampling case. The revised linear ephemeris of Feline et~al.
(2005) was based on a sparsely sampled (O$-$C) diagram. If the timings
of Wood et~al.  (1995), Mukai et~al. (1997) and Ioannou et~al. (1999)
were included in their diagram the orbital period modulation would
have become clear.

The middle panel of Fig.~\ref{fig:simul} illustrates the effect of the
time coverage on the detection of a period modulation. In this case
the observations cover only about half of the cycle period (solid
circles), leading to an incorrect inference of a long-term orbital
period decrease (solid curve). A longer baseline is needed to allow
identification of the cyclical nature of the period changes. Z~Cha is
an illustrative example of this case. Robinson et~al. (1995) inferred
a significant period increase from an (O$-$C) diagram covering 18~yr
of observations. Only when the time coverage was increased to 30~yr
the cyclical behavior of the period changes became clear (Baptista
et~al. 2002).

The lower panel of Fig.~\ref{fig:simul} shows how the accuracy of the
eclipse timings affect the ability to detect period modulations. In
this case the amplitude of the modulation was reduced to match the
larger uncertainty of the eclipse timings (20~s). The period
modulation is lost in the noise, despite the fact that the (O$-$C)
diagram has good sampling and time coverage (solid circles), and the
best-fit ephemeris is the linear one (solid line). FO~Aqr, with
inclination $i \sim 70^{\circ}$ and grazing eclipses (Hellier et~al.
1989), may be an example of this case. The large uncertainty of its
eclipse timings ($\sim 180$~s) is enough to mask cyclical period
modulations of amplitude similar to those seen in other eclipsing CVs.

In summary, in order to be able to detect a period modulation one
needs a well-sampled (one data point every 1$-$3~yr, no big gaps)
(O$-$C) diagram covering at least a decade of observations,
constructed from precise eclipse timings (uncertainty $\simlt 20$~s).

In order to construct a sample according to these requirements, we
searched the CVcat database\footnote{CVcat/TPP is a web-based
  interactive database on cataclysmic variable stars ({\tt
    http://cvcat.net/}).} for all eclipsing CVs with inclination $i
\geq 70^{\circ}$. The accuracy of eclipse timings below this limit is
not enough to allow detection of period modulations with amplitudes
$\simlt 200$~s. We find 14 eclipsing CVs satisfying the above
criteria, 6 systems below and 8 systems above the period gap. They are
listed in Table~\ref{tab:modcvs}. \emph{All systems in the sample show
  cyclical period changes}. With the inclusion of HT~Cas, there is
presently no CVs with well sampled and precise (O$-$C) diagram
covering more than a decade of observations that do not show cyclical
period changes. This underscores the conclusion of Baptista et~al.
(2003) that cyclical period changes seem a common phenomenon in CVs,
being present equally among systems above and below the period gap.

\begin{table}
  \caption{Observed orbital period modulations in CVs}
  \begin{tabular}{lcccl}
    \hline
    Object & $P_{{\rm orb}}$ & $P_{{\rm mod}}$ & $\Delta P/P$       & Ref.\\
           & (hr)            & (yr)            & ($\times 10^{-6}$) &     \\ 
    \hline
    V4140~Sgr & 1.47 &  6.9 & 0.93 & 1 \\ 
    V2051~Oph & 1.50 & 22.0 & 0.30 & 1 \\
    OY~Car    & 1.51 & 35.0 & 0.52 & 2 \\ 
    EX~Hya    & 1.64 & 17.5 & 0.55 & 3 \\
    HT~Cas    & 1.77 & 36.0 & 0.44 & This work.\\
    Z~Cha     & 1.79 & 28.0 & 0.85 & 4 \\
    IP~Peg    & 3.80 &  4.7 & 8.00 & 5 \\
    U~Gem     & 4.17 &  8.0 & 3.00 & 6 \\
    DQ~Her    & 4.65 & 13.7 & 2.00 & 7 \\
    UX~UMa    & 4.72 & 7.1,\,10.7,\,30.4$^{\ast}$ & 2.60 & 8 \\
    T~Aur     & 4.91 & 23.0 & 3.10 & 9 \\
    EX~Dra    & 5.03 &  5.0 & 7.90 & 10 \\
    RW~Tri    & 5.57 & 7.6,\,13.6$^{\ast}$ &2.10 & 11 \\  
    AC~Cnc    & 7.21 & 16.2 & 5.50 & 12 \\
    \hline
    \multicolumn{5}{l}{$\ ^{\ast}$Multiperiodic (the best determined
    periodicities are considered).} \\
    \multicolumn{5}{l}{\emph{References}: (1) Baptista et~al. 2003; (2) 
      Greenhill et~al. 2006;} \\
    \multicolumn{5}{l}{(3) Hellier \& Sproats 1992; (4) Baptista et~al. 
      2002; (5) Wolf et} \\
    \multicolumn{5}{l}{al. 1993; (6) Warner 1988; (7) Zhang et~al. 
      1995; (8) Rubenstein} \\
    \multicolumn{5}{l}{ et~al. 1991; (9) Beuermann \& Pakull 1984;
      (10) Shafter \&} \\
    \multicolumn{5}{l}{Holland 2003; (11) Robinson et~al. 1991;
    (12) Qian et al. 2007b.}
  \end{tabular}
  \label{tab:modcvs}
\end{table}

Apsidal motion is not a viable explanation for such period changes
because the orbital eccentricity for close binaries is negligible. The
presence of a third body in the system has often been invoked as an
alternative explanation. However, a light-time effect implies a
strictly periodic modulations in orbital period, which is usually not
observed when data covering several cycles of the modulation are
available. If one is to seek a common explanation for the orbital
period modulation seen in CVs, then all periodic effects (such as a
third body in the system) must be discarded, since the observed period
changes in several of the systems (e.g., UX~UMa, RW~Tri, V2051~Oph)
are cyclical but clearly not strictly periodic.

The best current explanation for the observed cyclical period
modulation is that it is the result of a solar-type magnetic activity
cycle in the secondary star. Amongst several mechanisms proposed to
explain such modulations, the hypothesis of Applegate (1992) seems the
most plausible. It relates the orbital period modulation to the
operation of a hydromagnetic dynamo in the convective zone of the
late-type component of close binaries.  More precisely, Applegate's
hypothesis assumes that a small fraction of the internal angular
momentum of the active component is cyclically exchanged between an
inner and an outer convective shell due to a varying internal magnetic
torque. This affects the oblateness and the gravitational quadrupole
moment of the active component, which oscillates around its mean
value. When the quadrupole moment is maximum, the companion star feels
a stronger gravitational force, so that it is forced to move closer
and faster around the centre of mass, thus attaining the minimum
orbital period.  On the other hand, when the quadrupole is minimum,
the orbital period exhibits its maximum. Lanza et~al. (1998) and Lanza
\& Rodon\`{o} (1999) have elaborated more on this idea. The model was
applied to a sample of CVs by Richman et~al. (1994). The fractional
period change $\Delta P/P$ is related to the amplitude $\Delta(O-C)$
and to the period $P_{{\rm mod}}$ of the modulation by (Applegate
1992),

\begin{equation}
\frac{\Delta P}{P}= 2\pi\,\frac{\Delta(O-C)}{P_{\rm mod}}= 4\pi\,\frac{A}{C}\; .
\label{eq:dpp}
\end{equation}

\noindent Using the values of $A$ and $C$ in Tables~\ref{tab:ephem}, we find
$\Delta P/P = 4.4 \times 10^{-7}$ for HT~Cas. The values of $\Delta
P/P$ for all the CVs in our sample are listed in the fourth column of
Table~\ref{tab:modcvs}.

The critical aspect of Applegate's hypothesis is the connection of his
model of gravitational quadrupole changes to a realistic cyclic dynamo
model capable of produce such modulations. In this regard, R\"{u}diger
et~al. (2002) presented an $\alpha^{2}$~dynamo model for RS~CVn stars,
adding a dynamo mechanism to Applegate's model. Also, theoretical
improvements and observational constrains appeared recently in the
literature in an attempt to overcome the limitations faced when
Applegate's model is applied to RS~CVn stars (Lanza \& Rodon\`{o}
2002, 2004; Lanza 2005, 2006a, 2006b). All these results can be scaled
to CVs above the period gap because their secondaries also have
convective envelopes.

On other hand, secondaries of CVs below the gap are thought to be
fully convective (i.e, masses $M_{2} < 0.3\,M_{\odot}$). Because fully
convective stars have no overshoot layer, the usual dynamo mechanisms
cannot be at work in these stars (Dobler 2005). However, isolated
late-type main sequence stars (spectral type M5 and later) show
indications of the presence of strong magnetic fields (Hawley 1993;
Baliunas et al.  1995; West et~al. 2004). Moreover, if magnetic
activity in the secondary star is used as an explanation for the
observed period modulations, the CVs below the period gap represent
the first sample of fully convective dwarfs with magnetic cycles
known. Alternative small-scale dynamo models have been proposed to
sustain magnetic fields and induce magnetic activity cycles in fully
convective stars (Durney et~al. 1993; Haugen et~al. 2004; Brandenburg
et~al. 2005; Dobler 2005 and references therein). These models
indicate that an overshoot layer is not a necessary ingredient for the
generation of large scale magnetic fields.

The common occurrence of $P_{{\rm orb}}$ modulations in CVs is
consistent with the results of Ak et~al. (2001). They found cyclical
variations in the quiescent magnitude and outburst interval of a
sample of CVs above and below the period gap, which they attributed to
solar-type magnetic activity cycles in the secondary stars. They also
found no correlation of the cycle period with the rotation regime of
the secondary star (i.e., orbital period, for the phase-locked
secondary stars in CVs). Considering the period modulations of a
variety of close binary systems, Lanza \& Rodon\`{o} (1999) found
similar observational evidence for system above and below the period
gap.

\subsection{A comparison of the observed orbital period modulations above and 
  below the period gap}
\label{sub:diff}

Given the sample of well observed CVs which exhibit cyclical orbital
period changes, this section attempts to quantify the common behavior
as well as to address the systematic differences between the observed
modulations in systems above and below the period gap. We extended the
comparison by considering the period modulations observed in other
longer-period close binaries (Algols, RS CVn and W UMa stars). Their
magnetic activity should resemble, in some sense, the magnetic
activity of the CVs above the gap since they also have a late-type
(active) component with a convective envelope.

\begin{figure}
  \includegraphics[scale=0.36]{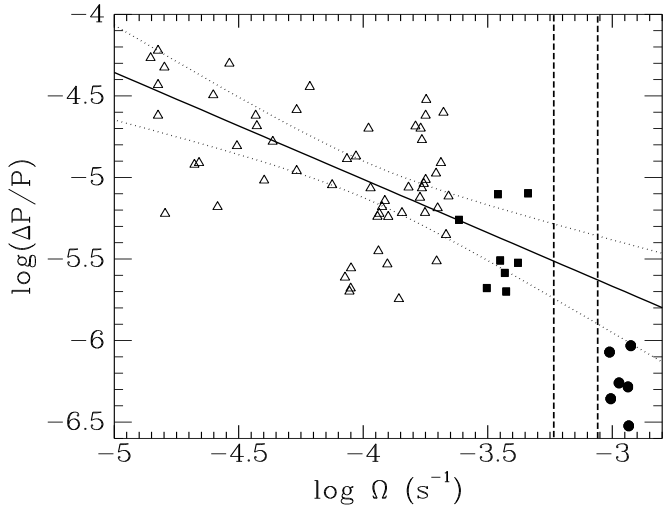}
  \caption{Diagram of the fractional period change $\Delta P/P$ 
    versus the angular velocity $\Omega$ of the active component star
    in short-period CVs (filled circles), long-period CVs (filled
    squares) and other close binaries from the literature (open
    triangles). The period gap is represented by the vertical dashed
    lines. The best-fit linear relation and its 3-$\sigma$ confidence
    level are shown as solid and dotted lines, respectively. The
    diagram includes the measurements of the CVs listed in
    Table~\ref{tab:modcvs}.}
  \label{fig:dppo}
\end{figure}

Figure~\ref{fig:dppo} shows a diagram of the fractional period change
$\Delta P/P$ versus the angular velocity $\Omega$ $= 2\pi/P_{{\rm
    orb}}$ of the active, late-type component star in close binaries
(rotation is a key ingredient of the dynamo action, see Lanza \&
Rodon\`{o} 1999). It includes data from the 14 eclipsing CVs listed in
Table~\ref{tab:modcvs}. The period gap is indicated by vertical dashed
lines. The open triangles depict $\Delta P/P$ values for other close
binaries with cyclical period modulations (Lanza \& Rodon\'{o} 1999;
Qian et al. 1999, 2000a, 2000b, 2002, 2004, 2005, 2007a; Lanza et al.
2001; Kang et al. 2002; Qian 2002a, 2002b, 2003; Yang \& Liu 2002,
2003a, 2003b; Zavala et al.  2002; \c{C}ak{\i}rl{\i} et al. 2003; Kim
et al. 2003; Qian \& Boonrucksar 2003; Af\c{s}ar et al. 2004; Lee et
al. 2004; Qian \& Yang 2004; Yang et al. 2004, 2007; Zhu et al. 2004;
Borkovits et al.  2005; Qian \& He 2005; Erdem et al. 2007; Pilecki et
al. 2007; Szalai et al.  2007).  Systems with independent evidence
that the observed (O$-$C) modulations can be explained by a third body
were excluded (actually, it is not possible to exclude the possibility
that the observed modulation of some of the systems plotted may still
be caused by a third body. However, for most of the systems the
observation of the modulation covers more than one cycle and there is
some indication that the variation is non-periodic). There is a clear
correlation between the fractional period change and the angular
velocity, which can be expressed by a relationship of the type $\Delta
P/P \propto \Omega^{\gamma}$ for all close binaries above the period
gap. A linear least-squares fit yields $\gamma = -0.7 \pm 0.1$.  The
best-fit solution and the 3-$\sigma$ confidence level are indicated by
a solid and dotted lines, respectively, in Fig.~\ref{fig:dppo}.

We find that the average fractional period changes of the short-period
CVs ($\overline{\Delta P/P} \simeq 8 \times 10^{-7}$) are
systematically smaller than those of the long-period CVs
($\overline{\Delta P/P} \simeq 4 \times 10^{-6}$) by a factor $\simeq
5$. The mean value of $\Delta P/P$ of the short-period CVs is more
than 3-$\sigma$ below the value expected from the linear relation
inferred for the other, longer period close binaries. An attempt to
simultaneously explain CVs above and below the gap by vertically
displacing the $\Delta P/P$ versus $\Omega$ fitted relation results in
a poor fit with high $\chi^{2}_{\nu}$. Similarly, including the CVs
below the period gap in the sample degrades the fit and significantly
increases the $\chi^{2}_{\nu}$. Despite the small sample (14 objects),
there is a statistically significant difference between the $\Delta
P/P$ values of the CVs above and below the period gap.  This
difference cannot be eliminated even if we take into account the
fitted $\Delta P/P$ versus $\Omega$ relation that predicts decreasing
$\Delta P/P$ values for increasing $\Omega$.  Thus, we are forced to
conclude that the CVs below the period gap do not fit the relation
valid for the binaries above the period gap.

If the interpretation of cyclical period changes as the consequence of
a solar-like magnetic activity cycle is correct, the existence of
cyclical period changes in binaries with fully convective active stars
is an indication that these stars do have magnetic fields, not only
capable of inducing strong chromospheric activity (e.g, Hawley 1993),
but also measurable magnetic activity cycles (Ak et~al.  2001; this
paper). On the other hand, the fact that the fractional period changes
of binaries with fully convective stars is systematically smaller than
those of stars with radiative cores is a likely indication that a
different mechanism is responsible to generate and sustain their
magnetic fields. In this regard, the observed lower $\Delta P/P$
values yield a useful constrain to any model that might be developed
to account for magnetic fields in fully convective stars.

The investigation of cyclical period changes in CVs will greatly
benefit from the increase in the (presently small) sample of systems
with well-sampled (O$-$C) diagrams covering more than a decade of
observations. This demands the patient but systematic collection of
precise eclipse timings over several years. It is worth mentioning
that there is a good number of well-known eclipsing CVs for which a
few years of additional eclipse timings observations would suffice to
overcome the ambiguities illustrated in Fig.~\ref{fig:simul} and to
allow statistically significant detection of period modulations. We
also remark that none of the systems \emph{inside} the period gap has
been yet observed for long enough time to allow identification of
cyclical period changes. It would be interesting to check whether
these systems will fit the $\Delta P/P \propto \Omega^{-0.7}$ relation
or will display a behavior similar to that of the CVs below the period
gap.

In order to explain the CV period gap, the disrupted braking model
predicts a significant reduction in magnetic braking efficiency
between the systems above and below the period gap (Hameury et~al.
1991). The observed difference in $\Delta P/P$ between CVs above and
below the period gap could be used to test this prediction if a
connection between fractional period changes and braking efficiency
could be established.

\begin{acknowledgements}
  B.W.B. acknowledges financial support from CNPq-MCT/Brazil graduate
  research fellowship. R.B. acknowledges financial suport from
  CNPq-MCT/Brazil throught grants 300.345/96-7 and 200.942/2005-0.
\end{acknowledgements}

\end{document}